\documentclass[a4paper,fleqn,usenatbib]{mnras}

\usepackage{mathptmx}
\usepackage[T1]{fontenc}
\usepackage{ae,aecompl}

\usepackage{graphicx}	% Including figure files
\usepackage{amsmath}	% Advanced maths commands
\usepackage{amssymb}	% Extra maths symbols

\newcommand{\msun}{${M}_{\odot}$}

\newcommand{\e}{$\pm$}
\newcommand{\ltsima} {$\; \buildrel < \over \sim \;$} 
\newcommand{\simlt}  {\lower.5ex\hbox{\ltsima}}            % < over MMM 
\newcommand{\gtsima} {$\; \buildrel > \over \sim \;$} 
\newcommand{\simgt}  {\lower.5ex\hbox{\gtsima}}            % > over MMM 
\title[Flickering of accreting white dwarfs]{Flickering of accreting  white dwarfs: the remarkable
amplitude - flux relation and disc viscosity}

\author[Zamanov et al.]{
R. K. Zamanov,$^{1}$\thanks{E-mail: rkz@astro.bas.bg, sboeva@astro.bas.bg, jeno@astro.columbia.edu, glatev@astro.bas.bg}
S. Boeva,$^{1}$ 
G. Latev,$^{1}$
J. L. Sokoloski,$^{2}$
K. A. Stoyanov,$^{1}$
\newauthor 
V. Genkov,$^{1}$ 
S. V. Tsvetkova,$^{1}$
T. Tomov,$^{3}$
A. Antov,$^{1}$
and  M. F. Bode$^{4}$ 
\vspace{0.3cm}
\\
$^{1}$Institute of Astronomy and National Astronomical Observatory, Bulgarian Academy of Sciences, Tsarigradsko Shose 72, \\
BG-1784 Sofia, Bulgaria
\vspace{0.1cm} \\
$^{2}$ Columbia Astrophysics Laboratory, Columbia University, 550 West 120th Street, New York, NY 10027, USA  
\vspace{0.1cm} \\
$^{3}$ Centre for Astronomy, Faculty of Physics, Astronomy and Informatics, Nicolaus Copernicus University, Grudziadzka 5, \\ 
87-100, Torun, Poland
\vspace{0.1cm} \\
$^{4}$ Astrophysics Research Institute, Liverpool John Moores University, IC2 Liverpool Science Park, Liverpool, L3 5RF, UK
}

\date{Accepted 2015 November 30. Received 2015 October 15; in original form 2015 October 15}

\pubyear{2015}

\begin{document}
%\label{firstpage}
\pagerange{\pageref{firstpage}--\pageref{lastpage}}
\maketitle

% Abstract of the paper
\begin{abstract}
We analyse optical photometric data of short term variability (flickering) 
of accreting white dwarfs in cataclysmic variables
(KR~Aur, MV~Lyr, V794~Aql, TT~Ari, V425~Cas), recurrent novae (RS~Oph and T~CrB) and jet-ejecting symbiotic stars
(CH~Cyg and MWC~560).
We find that the amplitude-flux relationship is visible over four orders of magnitude,
in  the range of fluxes from  $10^{29}$ to $10^{33}$ erg~s$^{-1}$~\AA$^{-1}$, as a
``statistically perfect" correlation with  correlation coefficient  0.96 and p-value $ \sim 10^{-28}$.
In the above range, the amplitude of variability for any of our 9 objects is proportional to the flux level 
with (almost) one and the same factor of proportionality for all 9 accreting white dwarfs 
with $\Delta F  =   0.36 (\pm  0.05) F_{av}$,  $\sigma_{rms} =  0.086(\pm 0.011)  F_{av}$, 
and $\sigma_{rms} / \Delta F = 0.24 \pm 0.02$. 
Over all, our results indicate that the viscosity in the accretion discs 
is practically the same  
for all 9 objects in our sample, in the mass accretion rate 
range  $2 \times 10^{-11} -  2\times10^{-7}$ $M_\odot$~yr$^{-1}$.
\end{abstract}

% Select between one and six entries from the list of approved keywords.
% Don't make up new ones.
\begin{keywords}
accretion, accretion discs -- (stars:) novae, cataclysmic variables  --  binaries: symbiotic 
%% -- stars: individual: RS Oph, T CrB, CH Cyg, MWC 560, KR Aur, TT Ari, V425 Cas, MV Lyr, V794 Aql
\end{keywords}

%%%%%%%%%%%%%%%%%%%%%%%%%%%%%%%%%%%%%%%%%%%%%%%%%%
%%%%%%%%%%%%%%%%% BODY OF PAPER %%%%%%%%%%%%%%%%%%

\section{Introduction}
Cataclysmic variables (CVs) are close binary stars 
consisting of a late-type main sequence star which is transferring material to the white dwarf. 
% that consist of   a white dwarf primary, and a 
%red dwarf 
% late-type main-sequence mass donor.    
% Interacting binaries of particular value are Cataclysmic Variables (CVs). These systems
% are composed of a white dwarf, accreting material via an accretion disc from a
% late-type main-sequence `secondary'.
% Cataclysmic variables are binary stars in which a relatively normal star is transferring mass to its compact companion. Thi
Symbiotic stars included here are wide binaries in
which material is transferred from an evolved red giant star to
a white dwarf.  

Flickering is one of the most intriguing characteristics of the accreting compact objects. 
It appears as broad-band stochastic light variations on time-scales of a few minutes 
with amplitude  from a few $\times 0.01$ mag to more than one magnitude.
Random fluctuations of the brightness are
observed throughout diverse classes of objects 
that accrete material onto a compact object (white dwarf, neutron star or black hole)  
% -- cataclysmic variables, symbiotic stars and supersoft X-ray binaries.
 -- binary stars, X-ray binaries, Active Galactic Nuclei.  
The source of the flickering variations is the accretion disk - either the disk itself or 
some parts of the disk, e.g. bright spot or boundary layer.
The first reported detection of flickering activity is by Pogson (1857) based on visual observations of the dwarf nova U~Gem.
Photoelectric observations identified the flickering as a common characteristic of the accretion process 
(e.g.  Mumford 1966, Henize 1949, Robinson 1973). 
% Bruch (1992) has made a systematic observational study of flickering in CVs.
A quantitative study of the flickering
properties in cataclysmic variables has been performed  by Bruch (1992), who defined several physical parameters 
to describe the phenomenon. 

% Who used the word flickering to describe the short term light variations?
% (to burn or shine fitfully or with a fluctuating light <a candle flickering in the window>) 
% (to burn unsteadily or with a constantly changing light <a flickering candle> ) 
% flickering (shining unsteadily)
% Random fluctuations have an amplitude of 0.1 - 1.0 mag at B and occur on timescales of 1 - 10 minutes  (1 hour or less). 
% rapid variability (= flickering) is cleary seen (maximum amplitude  ?  mag in B).

The amplitude-flux relation has previously been discovered for a few objects on an object-by-object basis. 
Here we present data for  9 accreting  white dwarfs showing flickering in the optical bands.
Our aim is to investigate the behaviour of the flickering amplitude 
and rms flux relative to the average flux of the hot component, 
examining the position of different objects on two diagrams: $\Delta F$ versus  $F_{av}$
and  $\sigma_{rms}$ versus  $F_{av}$.

% In this paper, we search for evidence of connection between the flickering and the flux of the accreting 
% white dwarf in the light curves of CVs, RNe, jet ejecting symbiotic stars. 
% We find good evidence for correlation between the brightness fluctuations and average flux of the hot component.

%-------------------------------------------------------------------------   
 \begin{figure*}    
   \vspace{3.5cm}     
   \includegraphics{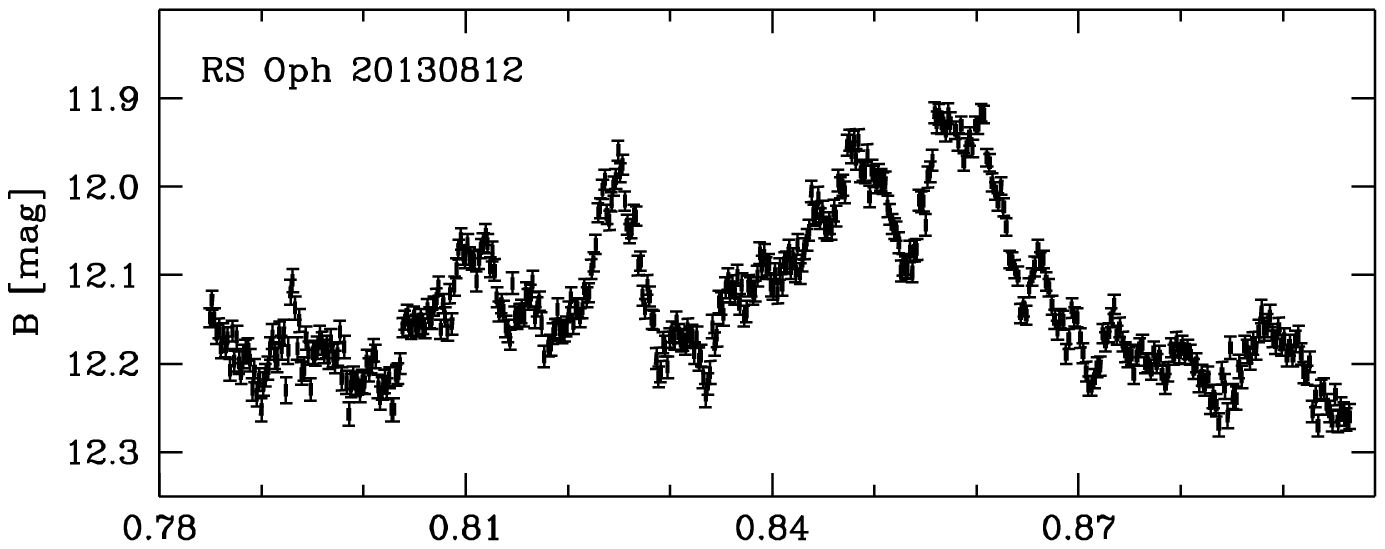}  
   \includegraphics{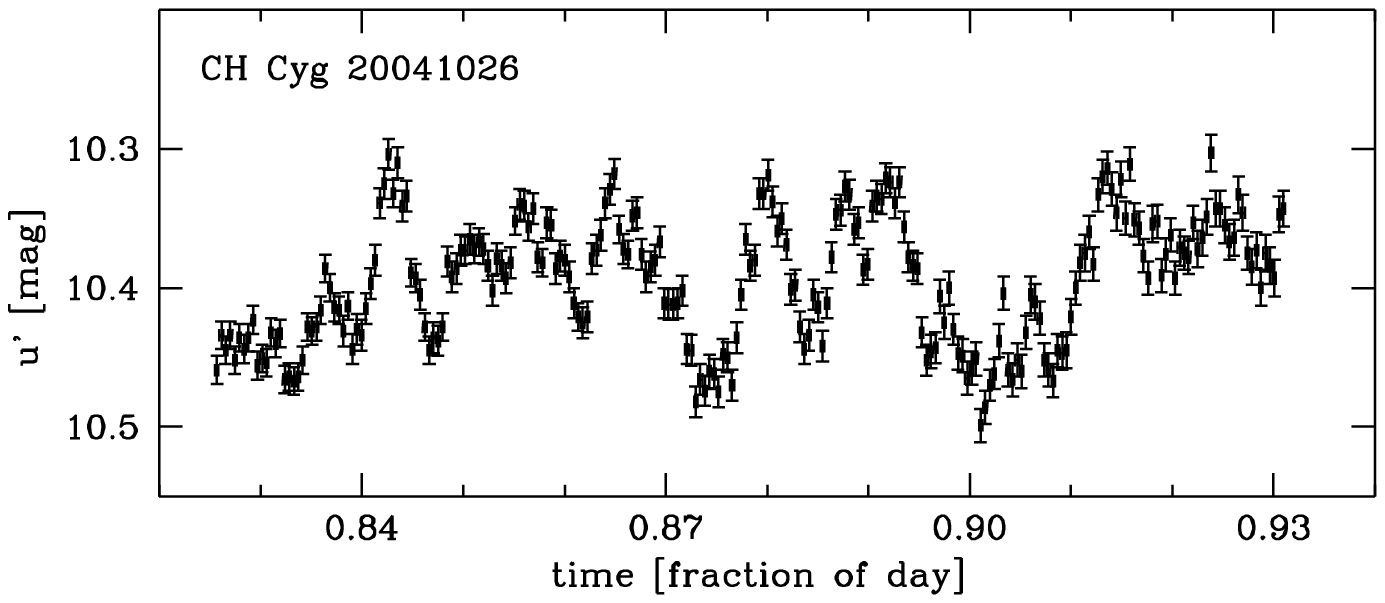}  
   \includegraphics{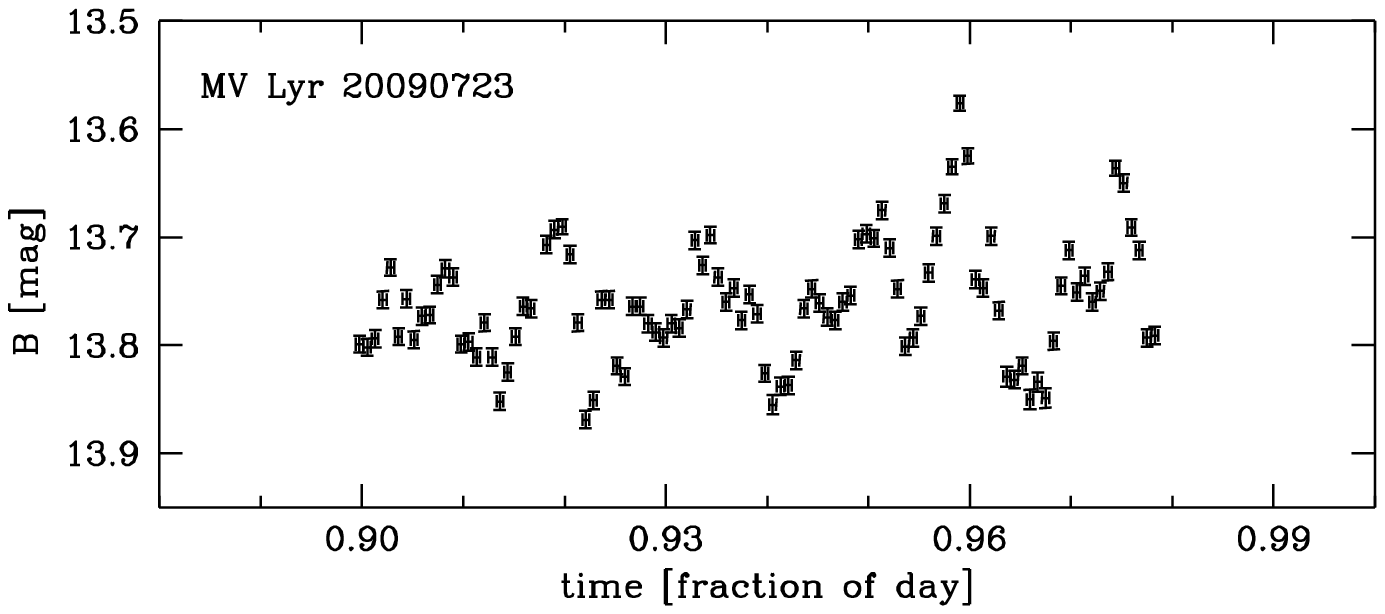}  
  \caption[]{Example light curves of RS~Oph, CH Cyg and MV~Lyr. The date of observations 
  (in format yyyymmdd) is indicated in each panel.  }
   \label{fig.example}      
 \end{figure*}	     
%------------------------------------------------------------------------------ 

%----------------------------------------------------
\begin{table*}
\centering
\caption{In the Table are given 
the name of the object, 
its type, 
$N_{obs}$ (the number of light curves), 
total duration of the observations in hours,
the brightness interval during our observations, 
the adopted distance in parsecs, the interstellar extinction, 
the adopted spectral type and visual  band magnitude ($m_B$) of the mass donor,  and finally 
the ratios amplitude/flux and rms/flux.
}  %, minimum and maximum brightness in B band. & B (min - max) }
\label{tab.obj}
\begin{tabular}{lllrrlrlllllll}
 \hline
  Object & type   & $P_{orb}$ & $N_{obs}$  &  D[h] &  min -- max	 & $d$ [pc] & $E_{B-V}$ & donor &  $\Delta F/ F_{av}$ &  $\sigma_{rms} / F_{av}$  & \\ 
 \hline
 T~CrB    & RecN    & 227 d   & 31 & 46.6  & $m_U$ 10.5-11.9  & 960      &  0.14	& M4III,   $m_B\approx$11.8   &  0.397\e0.109 & 0.083\e0.023   & \\
 RS Oph   & RecN    & 455 d   & 76 & 139.7 & $m_B$ 11.1-13.2  & 1600     &  0.73	& M2III,   $m_B\approx$13.9   &  0.349\e0.103 & 0.086\e0.036   & \\
 MWC~560  & symbio  & 1931 d  & 21 & 46.6  & $m_U$  9.3-11.5  & 2500     &  0.15	& M5.5III, $m_B\approx$13.6   &  0.326\e0.130 & 0.077\e0.032   & \\
 CH~Cyg   & symbio  & 15.6 yr & 16 & 26.1  & $m_U$  7.3-10.8  & 244      &  0.20	& M6III,   $m_B\approx$11.5   &  0.424\e0.160 & 0.093\e0.041   & \\
  \\  
 KR~Aur   & CV      & 3.91 h  & 12 & 42.2  & $m_B$  13.1-18.9 & 1000     &  0.05	& M1V,     $m_B\approx$21.5   &  1.248\e1.000 & 0.306\e0.247   & \\ 
 V425 Cas & CV      & 3.59 h  & 14 & 17.3  & $m_B$  14.7-15.7 &  700     &  0.30	& M3V,     $m_B\approx$18.5   &  0.326\e0.161 & 0.089\e0.047   & \\
 MV~Lyr   & CV      & 3.19 h  & 20 & 35.9  & $m_B$  12.8-14.6 &  505     &  0.00	& M4V,     $m_B\approx$17.9   &  0.345\e0.086 & 0.079\e0.022   & \\ 
 V794~Aql & CV      & 3.68 h  &  6 &  9.6  & $m_B$  15.1-16.5 &  690     &  0.20	& M1V,     $m_B\approx$20.0   &  0.331\e0.079 & 0.088\e0.019   & \\  
 TT~Ari   & CV      & 3.30 h  &  8 & 32.0  & $m_B$  10.4-10.9 &  335     &  0.05	& M3.5V,   $m_B\approx$18.5   &  0.318\e0.073 & 0.069\e0.015   & \\  
 \hline  
 \end{tabular}  
 % \bottomrule
\end{table*}
%----------------------------------------------------

%-------------------------------------------------------------------------   
 \begin{figure}    
   \vspace{6.0cm}     
   \includegraphics{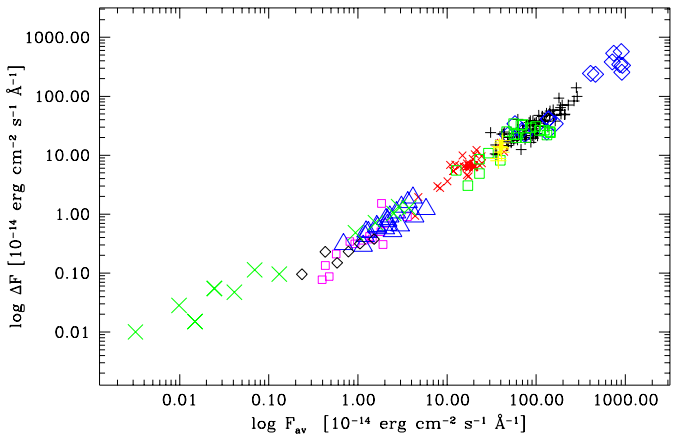}  
   \caption[]{Amplitude of the flickering  versus the average flux of the hot component, on a logarithmic scale
     for nine accreting white dwarfs (see Section 3 for details of the objects plotted). 
  %  RS~Oph -- black plusses, T~CrB -- red crosses, MWC~560 -- green squares,  CH~Cyg - blue diamonds, 
  %  V425~Cas -- magenta squares, KR Aur -- green crosses, MV~Lyr -- blue triangles, 
  %  V794~Aql - black diamonds, TT~Ari - yellow crosses. 
    Estimated errors are less than or equal to  the size of the symbols on the plot. 
    Remarkably, above $2 \times 10^{-15}$~erg~cm$^{-2}$~s$^{-1}$~\AA$^{-1}$, 
    all data points  lie on the same line with Y and X increasing/decreasing
    together.}
   \label{fig.1}      
 \end{figure}	     
%------------------------------------------------------------------------------ 

\section{Observations and data analysis}

We started to observe flickering of accreting white dwarfs in 1990. 
Over $\approx 25$ years of observations, we acquired
photometry of rapid variability
with the 2.0m RCC, 50/70~cm Schmidt and 60~cm telescopes of the  National Astronomical Observatory Rozhen,
the 60 cm telescope of the Belogradchick Astronomical Observatory (Bulgaria),
1.0-meter Nickel  telescope at UCO/Lick Observatory on Mt. Hamilton near San Jose, CA (USA), 
and  the fully robotic 2.0~m Liverpool Telescope\footnote{The Liverpool Telescope is operated on the island of La Palma by Liverpool John Moores University in the Spanish Observatorio del Roque de los Muchachos of the Instituto de Astrofisica de Canarias with financial support from the UK Science and Technology Facilities Council.} \citep{2004SPIE.5489..679S}. 

We have obtained more that 396 hours (204 light curves)  in total of observations of the flickering 
of the recurrent novae  RS~Oph and T~CrB; of the jet ejecting symbiotic stars MWC~560 and CH~Cyg, and 
of the CVs  KR~Aur, MV~Lyr, V425~Cas, V794~Aql and TT~Ari. 
In each run, brightness fluctuations on a timescale of $\sim 10$   %  5-15 minutes 
minutes are clearly visible.
% We find good evidence for small-amplitude brightness fluctuations on a timescale of 10 minutes or less. 
 
To analyze brightness fluctuations, we begin by
a conversion of the magnitudes into fluxes, adopting   
the calibration for a zero magnitude star of Bessel (1979).  
The observed flux during a given night was corrected   
for the contribution of the mass donor (red giant in case of symbiotic stars and 
red dwarf in case of CVs) and interstellar extinction. 
For the correction of the  mass donor contribution 
we adopt B band magnitudes and spectral types given in Table~1, and 
(U-B) and (B-V) colours for the corresponding spectral type given by Schmidt-Kaler (1982).

For each run, we calculate the  following dereddened quantities: 
$F_{max}$ -- the maximum flux of the hot component; 
$F_{min}$ -- the minimum flux of the hot component; 
$\Delta F = F_{max} - F_{min} $ -- peak-to-peak amplitude of the flickering; 
$F_{av}$ -- the average flux of the hot component:    
\begin{equation}
  F_{av} = \frac{1}{N}  \sum_{i=1}^{N} F_i; 
\end{equation}
and the absolute rms amplitude of variability  (the square-root of the light-curve variance):
\begin{equation}
  \sigma_{rms} =  \sqrt{ \frac{1}{N-1}  \sum_{i=1}^{N} (F_i - F_{av} )^2  }.
\end{equation}
Subsequently, we subtract the contribution expected from measurement errors.
% in a way similar to Eq.8 of Vaughan Edelson, Warwick and  Uttley  2003 MNRAS   as given in .
The corrections of $\Delta F$ and  $\sigma_{rms}$ for these measurement errors are small, in the range 1--4 per cent.  

Our runs have durations from 21 to 468 minutes (typical duration is about 100 minutes),  
the number of the points in one run is between 17 and 2400 (typically about 200 points) and 
the exposure time ranges from 1 to 300 seconds. A few examples of our observations are presented in
Fig.~\ref{fig.example}. 

% Here we do show what is really visible in our observations without cutting, filling the gaps,
% r any other manipulation.  

%%%%-----------------------------------------------------------------------------------------

\section{Results}

\begin{figure}    
   \vspace{9.5cm}     
   \includegraphics{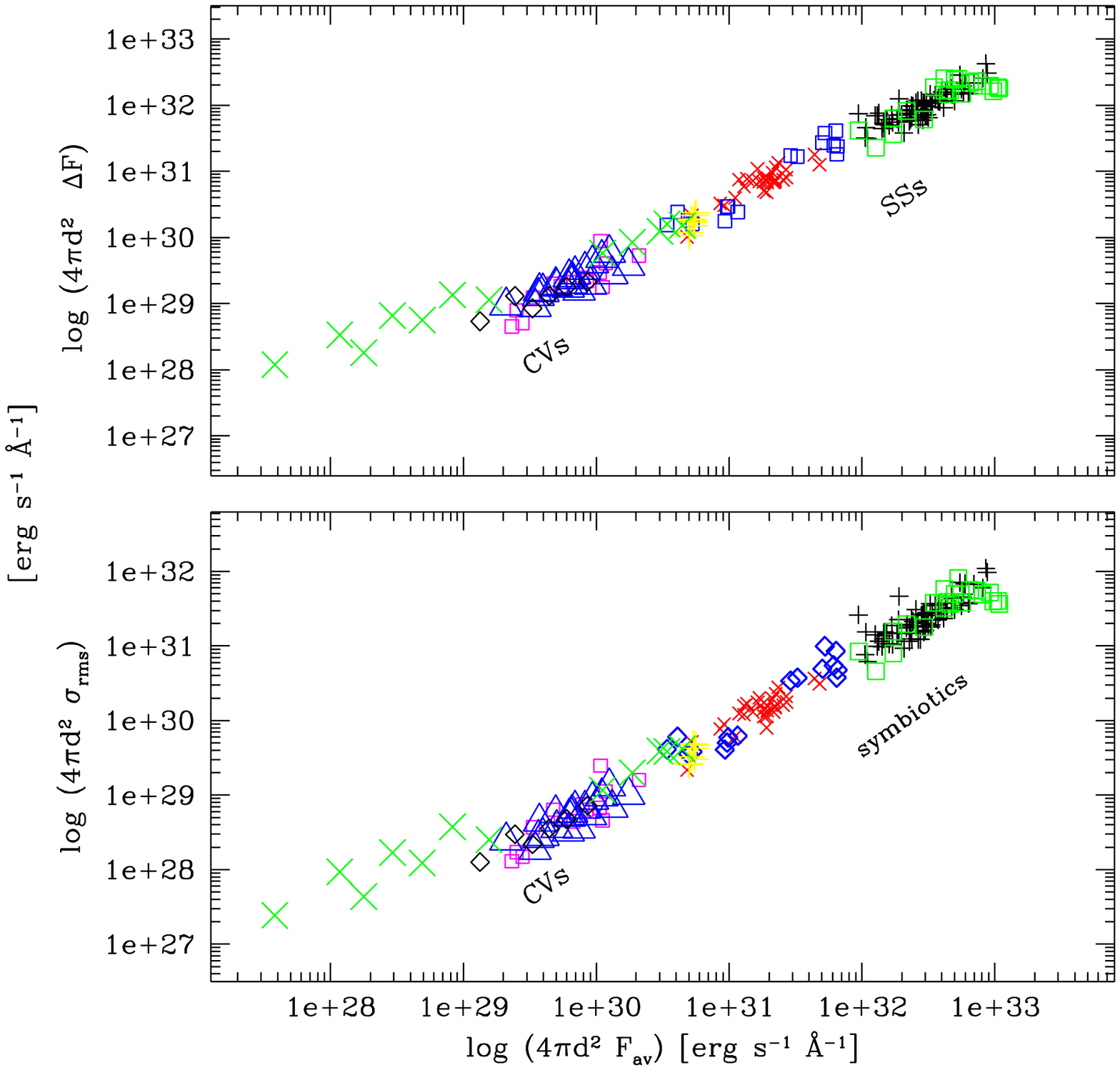}  
   \caption[]{Amplitude ($\Delta F$ -- top panel) and rms flux ($\sigma_{rms}$) versus the average flux of the hot component
    on a logarithmic scale corrected for the distance and interstellar extinction. 
    The symbols and the errors are the same as in Fig.\ref{fig.1}.
    In the X-axis range from $10^{29}$ to $10^{33}$ erg~s$^{-1}$~\AA$^{-1}$  (almost) all 
    data points lie  on one straight line. }
   \label{fig.mi}      
% \end{figure}	     
%------------------------------------------------------------------------------ 
%\newpage 
%\clearpage
%-----------------------------------------------------------------------------   
% \begin{figure}    
   \vspace{9.5cm}     
   \includegraphics{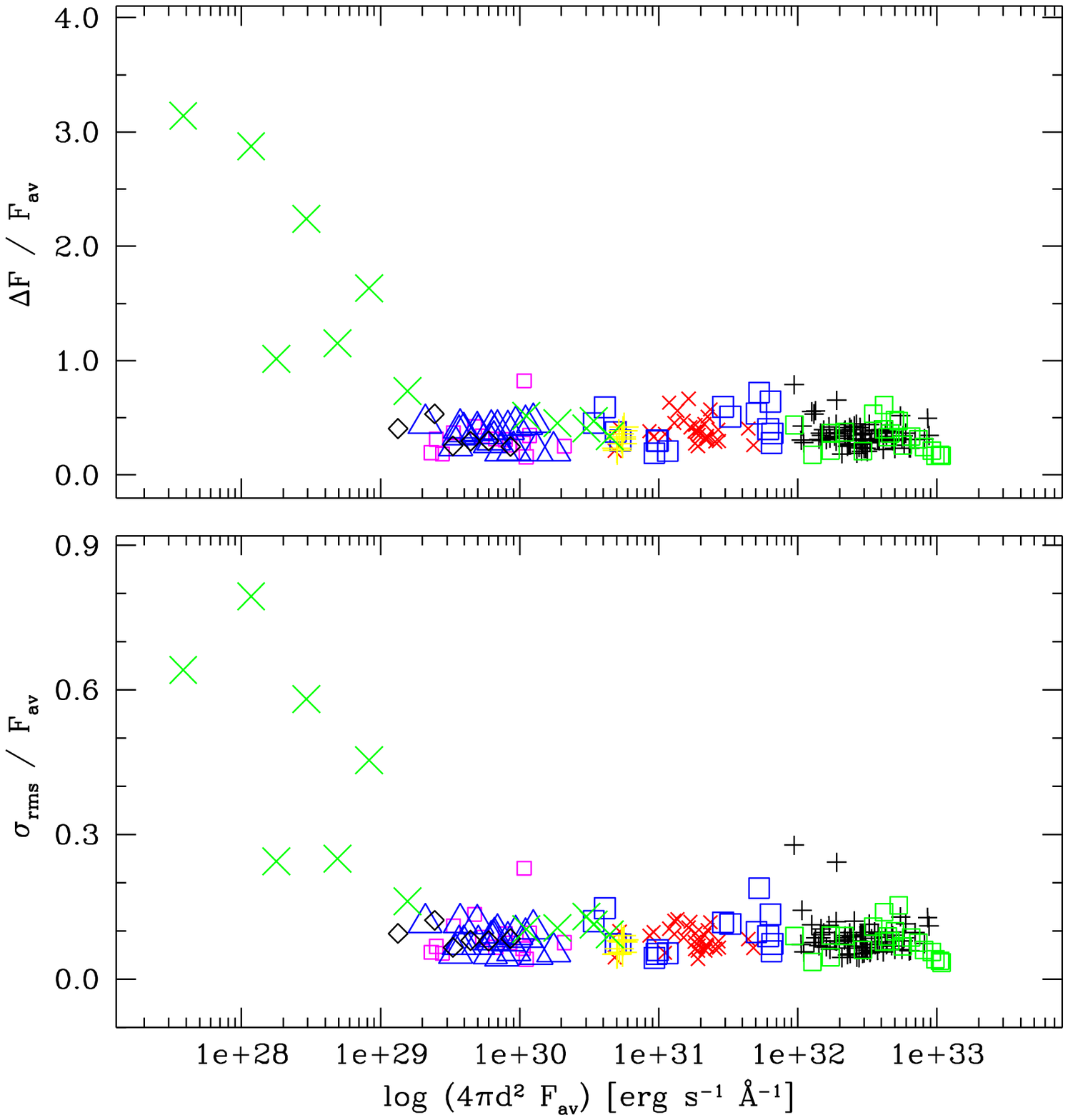}  
   \caption[]{Normalized amplitude ($\Delta F / F_{av}$ -- top panel) and normalized 
    rms  ($\sigma_{rms} / F_{av}$) versus the average flux of the hot component.  
    This figure is similar to Fig.\ref{fig.mi}, however the normalized quantities are plotted. }
   \label{fig.norm}      
 \end{figure}

%{\bf Fig.\ref{fig.1} :::  } 
In Fig.\ref{fig.1} we plot the  amplitude of the flickering  versus the average flux of the hot component,
on a logarithmic scale. 
These are the fluxes as observed on the Earth, corrected for the interstellar extinction. 
% and contribution of the  mass donor.
Each object is plotted with a different symbol:
    RS~Oph -- black plusses, 
    T~CrB -- red crosses, 
    MWC~560 -- green squares, 
    CH~Cyg - blue diamonds, 
    V425~Cas -- magenta squares, 
    KR Aur -- green crosses, 
    MV~Lyr -- blue triangles,       % MV~Lyr  is member of VY Scl subclass of NLs,
    V794~Aql - black diamonds, 
    TT~Ari - yellow crosses.        % TT Ari is a nova-like cataclysmic variable belonging to  VY Scl  group
The symbols used are the same in Fig.\ref{fig.1}, Fig.\ref{fig.mi} and Fig.\ref{fig.norm}. 

It is clearly apparent that for F$_{\rm av} \ge  2 \times 10^{-15}$~erg~cm$^{-2}$~s$^{-1}$~\AA$^{-1}$ 
all  198 data points are located  on a straight line with Y and X increasing/decreasing  together.

The quantities corrected for the distance are plotted in Fig.\ref{fig.mi}.  
In this figure we plot the amplitude of the flickering ($\Delta F$, upper panel)
and  the rms ($\sigma_{rms}$, lower panel) versus the average flux of the hot component.
All the quantities are corrected for the distance using the distances given in Table~1. 
In Fig.\ref{fig.mi}, the objects located 
above $\approx 10^{31}$ erg~s$^{-1}$~\AA$^{-1}$ are symbiotic stars, and 
below this value are CVs. This  reflects the fact that 
in symbiotics the mass donor is a red giant star and it is able to transfer more material 
than a red dwarf mass donor in a CV.

A comparison between Fig.\ref{fig.1} and the upper panel of Fig.\ref{fig.mi}  
shows that the objects change their places and the relationship remains the same. 
For example CH~Cyg, which is the closest object to the Earth in our sample, having a distance of only 244~pc,
is located in the upper right corner on Fig.\ref{fig.1}. 
When the distance is taken into account (Fig.\ref{fig.mi}), the recurrent nova RS~Oph is placed in the upper right corner
together with the jet-ejecting symbiotic MWC~560 -- 
the objects having the highest mass accretion rates ($\sim 10^{-7}$ \msun~yr$^{-1}$) among those in our sample.

In  Fig.\ref{fig.mi} it is seen that, when 
the quantity  $4 \pi d^2 F_{av}$  is  in the range  of  $10^{29} - 10^{33}$~erg~s$^{-1}$~\AA$^{-1}$, 
all the objects lie on one straight line. 
Mathematical tests show that we have a ``statistically perfect" correlation 
with Pearson correlation coefficient  0.99,  Spearman's (rho) rank correlation 0.98, and significance $p-value \approx 10^{-40}$. 
We fit the data to a straight line in log-log space (Fig.~\ref{fig.mi}), taking errors into account 
and obtain: 
\begin{eqnarray}
  \log (4 \pi d^2 \Delta F)     =0.996(\pm 0.043) \log (4 \pi d^2 F_{av}) -0.33(\pm 0.68)    \\
  \log (4 \pi d^2 \sigma_{rms}) =0.994(\pm 0.042) \log (4 \pi d^2 F_{av}) -1.10(\pm 0.44)  
\end{eqnarray} 
Thus, the relationship between the amplitude of variability  and the 
average flux of the hot component is consistent with linearity  for our 198 points in the 
range  $10^{29} - 10^{33}$~erg~s$^{-1}$~\AA$^{-1}$.

Our data are not evenly distributed among different objects, e.g. for RS~Oph we have more than 70 light curves, while for 
V794~Aql only 6. To check the influence of this distribution, 
we  applied bootstrap resampling (e.g. Davison \& Hinkley, 1997), 
selecting on a random basis 5 observations per object and repeating it $\sim 100$ times. 
For each sample  containing 45 points we recalculated the correlation and the result is always good, 
with correlation coefficient  $>0.95$ and significance better than $10^{-28}$. We repeated this procedure with 
6 observations per object and the values are similar, thus confirming the result based on all data.

In Fig.\ref{fig.norm} we plot the normalized quantities 
$\Delta F / F_{av}$ and $\sigma_{rms} / F_{av}$. Most of the values 
are in the range  $ 0.04  < \sigma_{rms} / F_{av} < 0.13$.
There are a few points that are considerably above the average. 
The deviating points are:  V425~Cas (20090723),  RS~Oph (20120815) and KR~Aur (January - February 2009). 
The record in our sample is the cataclysmic variable  KR~Aur, which 
in a low state achieves values   $m_V \approx 18.7$~mag, flickering amplitude 
1.2 mag,  
% ($V \sim 18.5$  20090120   20090226) achieves values 
$\Delta F / F_{av} \sim 1.5$ and   $\sigma_{rms} / F_{av} \sim 0.4$. 
When KR~Aur is brighter than $m_B \sim 16.5$~mag its flickering is similar to that of the other objects and follows
the same straight line. 

% Over the range  $10^{29} <  4 \pi d^2  F_{av} < 10^{31}$~erg~s$^{-1}$~\AA$^{-1}$
Excluding KR Aur in low state, 
we calculate mean  values  
$\Delta F / F_{av} \approx  0.362 \pm  0.045$,   
$\sigma_{rms} / F_{av} \approx  0.086 \pm 0.011$,
and  $\Delta F / \sigma_{rms} \approx  4.2 \pm 0.4$.  
Our results imply that the normalized amplitude ($\Delta F / F_{av}$), 
the normalized rms variability  ($\sigma_{rms} \ / F_{av}$), and the ratio  $\sigma_{rms}/ \Delta F$ 
are approximately independent of the source brightness over the range 
$10^{29} <  4 \pi d^2  F_{av} < 10^{31}$~erg~s$^{-1}$~\AA$^{-1}$ (see Fig.\ref{fig.norm}).

%-------------------------------------------------------------------------   
%------------------------------------------------------------------------------ 

\section{Discussion}
It is known that most of the accretion-powered sources exhibit random fluctuations in their flux.
A fundamental characteristic of fast stochastic variability is the correlation 
between variability amplitude and average flux.  %  (so called rms-flux relation). 
This relation is valid over a wide-range of timescales. 
Uttley, McHardy, \& Vaughan (2005) studied non-linear X-ray variability of X-ray binaries and active galaxies
and found a linear relation between rms and flux calculated from light curve segments.
The detection of this relation is reported for the galactic black hole binary Cyg~X-1 
and in the  accreting millisecond pulsar SAX J1808.4-3658 (Uttley \& McHardy 2001), 
in the ultra-luminous X-ray source  NGC 5408 X-1 (Heil \& Vaughan 2010),
in the extreme narrow-line Seyfert 1 galaxy IRAS 13224-3809 (Gaskell 2004)
and in the bright Seyfert 1 galaxy Markarian 766 (Vaughan et al. 2003). 
This feature of the broad-band  X-ray variability of accreting black holes
in X-ray binaries  and Active Galactic Nuclei 
is called the rms-flux relation.
The light curves, obtained with the $Kepler$ satellite, 
show the same linear relation in the case of the cataclysmic variables 
MV Lyr (Scaringi et al. 2012), V1504~Cyg, and KIC8751494 (Van de Sande, Scaringi, \& Knigge 2015).  
This relationship was also detected in  the recurrent nova RS~Oph (Zamanov et al. 2015).
The detection of the rms-flux relation in such different objects 
is a demonstration of
the universal nature of accretion-induced variability (Scaringi et al. 2012).
Here, we report the detection of a linear relationship between the root mean square (rms) variability
amplitude and the mean flux, which is valid not only for a single object but for 9 objects 
in accreting white dwarf binary systems. 
This relationship represents proof that the sources become more variable as they get brighter. 
The average flux of the hot component should be proportional to the mass accretion rate: 
$ 4 \pi d^2 F_{av} \propto M_{wd} \;  R_{wd}^{-1} \; \dot M_{acc} $.  
When the mass accretion rate increases, the average flux also increases, 
and the amplitude of the flickering also increases. 
The observations reported here indicate that the amplitude-flux relationship is valid over 
four orders of mass accretion rate, from about 
$\sim 2 \times 10^{-11}$ $M_\odot$~yr$^{-1}$ to $\sim 2\times10^{-7}$ $M_\odot$~yr$^{-1}$.
The amplitude of variability at any given moment and for any of our 9 objects is proportional to the flux level 
with a very similar factor of proportionality for all nine accreting white dwarfs
$\Delta F  =   0.35 (\pm  0.10) F_{av}$ and  $\sigma_{rms} =  0.08(\pm 0.03)  F_{av}$.  
In the range $10^{29} <  F_{av} \le 10^{31}$~erg~s$^{-1}$~\AA$^{-1}$, we have only 4 deviating points from 198 runs. 
This indicates that deviations from the rms-flux relationship do exist, but they are relatively rare, occurring in $\sim 2$ per cent 
of the cases. 

From Fig.\ref{fig.mi} and Fig.\ref{fig.norm} it seems that 
the above relationship is not valid, when 
$F_{av}$ is below $10^{29}$~erg~s$^{-1}$~\AA$^{-1}$. 
This flux corresponds approximately to a mass accretion rate 
$\approx 2 \times 10^{-11}$ $M_\odot$~yr$^{-1}$. 
It might be connected with a critical mass accretion rate below which the disc structure changes. 
Because this suspicion is based only on
one object (KR~Aur), more data for low states of CVs would be helpful to determine
where are the exact limits of validity of the linear rms-flux relation. 

The broadband variability is often attributed to inward propagating fluctuations driven by
stochasticity in the angular momentum transport mechanism (Lyubarski 1997).
Cowperthwaite  \& Reynolds (2014) presented a non-linear numerical model for a geometrically thin accretion disk
with the addition of stochastic non-linear fluctuations in the viscous parameter, 
capable of reproducing  the observed linear rms-flux relationship in the disk luminosity. 
King et al. (2004)  have found that the normalized rms variability is roughly a constant for each value of the viscosity parameter $\alpha$. 
Following  this, our results seem to indicate that the viscosity in the accretion disks 
($\alpha$) is almost identical
for all 9 objects in our sample, in the mass accretion rate 
range  $2 \times 10^{-11} -  2\times10^{-7}$ $M_\odot$~yr$^{-1}$.

\section{Conclusions}
On the basis of 204 light curves of flickering of nine accreting white dwarfs in CVs and symbiotic stars,  
we calculated the amplitude and rms of variability. 
We report a remarkably linear relation between the peak-to-peak amplitude 
(and rms variability) and the mean flux  holding in the range 
$10^{29} - 10^{33}$ erg~s$^{-1}$~\AA$^{-1}$.
In this range all objects follow practically the same relation. 
The amplitude - flux ($\Delta F$ versus $F_{av}$) and rms  -  flux ($\sigma_{rms}$ versus $F_{av}$)  relationships
contain information about the dynamics of the infalling matter 
and are likely to occur in many more accreting systems.

%%%%%%%%%%%%%%%%%%%%%%%%%%%%%%%%%%%%%%%%%%%%%%%%%%

% \vskip 10.0cm 
% \newpage
% \clearpage

% Don't change these lines
 \bsp	% typesetting comment
 \label{lastpage}

\end{document}